# Effects of Repetitive Low-Level Blast Exposure on Mental Health in Veterans

Albert Zeng, BA[a*] (alzeng@upenn.edu) ; Iris Horng, MA[a] (ihorng@upenn.edu) David Park, BA[b] (davidparkjep@gmail.com); Nic Gray[c] (nic.gray@usjag.org); Randel Swanson, DO, PhD[d,e] (Randel.Swanson@upenn.edu); Dylan S. Small, PhD[a] (dsmall@upenn.edu)

a. Department of Statistics and Data Science, University of Pennsylvania, 265 S. 37th Street, Philadelphia, PA, 19104, USA

b. Department of Neuroscience, University of Pennsylvania, 415 Curie Blvd, Philadelphia, PA, 19104, USA

c. Uniformed Services Justice & Advocacy Group, 924 N Wahsatch Ave

Colorado Springs, PA, 80903, USA

d. Department of Physical Medicine and Rehabilitation, University of Pennsylvania, Perelman School of Medicine, 3400 Civic Center Blvd, Philadelphia, PA, 19104, USA

e. Center for Neurotrauma, Neurodegeneration and Restoration, Corporal Michael J. Crescenz Virginia Medical Center, 3900 Woodland Ave, Philadelphia, PA, 19104, USA

*Corresponding Author

**Abstract:**

Links between blast exposure and mild traumatic brain injury (mTBI), as well as other negative physical and mental effects are well established. However, the effects of low-level blast exposure, which does not rise to the level of a mTBI diagnosis, are less well-documented. Low-level blast exposure may initiate or accelerate neurodegenerative changes, leading to accelerated age-related cognitive decline or overt neuropathological diagnoses such as Chronic Traumatic Encephalopathy (CTE), in addition to other non-neurodegenerative long-term harmful effects, such as clinical depression. This study analyzed the effect of low-level blast exposure through a survey of military veterans who served in the U.S. military as mortarmen (n = 42). Survey participants responded to screening assessments measuring depression, PTSD, anxiety, and alcohol use disorder. Through a matched pair study design, surveyed mortarmen were compared to two control groups consisting of veterans without a history of mTBI exposure (n = 283) and veterans who had suffered at least one mTBI exposure (n = 1264). We found that surveyed mortarmen reported statistically significant higher depression (as measured by PHQ-9) than both groups of controls. Even after accounting for Post-Traumatic Stress disorder (PTSD) and unhealthy alcohol use, mortarmen still reported higher depression, suggesting that PTSD was not the sole factor responsible for worse mental health symptoms. These results highlight the importance of more comprehensive study of cumulative blast overpressure and its relationship with mental health.

# Introduction:

Mild traumatic brain injury (mTBI) has been described as the signature wound of the Iraq and Afghanistan Wars.(Lindquist et al., 2017) Samples of military veterans from Iraq and Afghanistan have found that between 15-20% of veterans suffered at least one deployment-related mTBI exposure, defined by a Glasgow Coma Scale score between 13 and 15 at 30 minutes post-injury as well as at least one of several symptoms.(Hoge et al., 2008; Lefevre-Dognin et al., 2021; Lindquist et al., 2017) The most common etiology of deployment-related mTBI diagnosis involves blast overpressure exposure(Lindquist et al., 2017) (most often with a secondary head impact).("Management and Rehabilition of Post-Acute mTBI," n.d.) As a result, particular emphasis has been placed on the diagnosis and analysis of blast-related mTBI.(Elder et al., 2014)

However, research on low-level blast exposure, which falls below the level of an mTBI, is limited in comparison.(Belanger et al., 2020; Belding et al., 2021; Elder et al., 2014) Low-level blast exposure is of particular concern as an occupational risk, as groups such as mortarmen and breachers experience low-level blast overpressure exposures regularly.(Kamimori et al., 2017; Stone et al., 2020) Existing safety regulations may not be sufficient for protecting against brain injury. For example, the mortar firing safety limit of 4 PSI overpressure is designed to protect against eardrum injury and not brain injury.(Kamimori et al., 2017) Research has shown that people occupationally exposed to blast exposure experience altered neurocognition in the immediate aftermath of high cumulative blast exposure, suggesting a possible link to sub-concussive injury.(Woodall et al., 2023) Repetitive low-level blast overpressure exposure is also linked to many long-term neurological effects.(Ahmed et al., 2013; Bryden et al., 2019; Finlay et al., 2012) Post-mortem analysis of the brains of blast-exposed veterans who served in Afghanistan has found evidence of Chronic Traumatic Encephalopathy.(Goldstein et al., 2012) Blast exposure is also associated with chronic mental health symptoms, including PTSD, depression, and neurobehavioral symptoms, in addition to other neurodegenerative effects, such as reduced white matter integrity and Alzheimer's risk.(Martindale et al., 2021; Trotter et al., 2015) Despite the likely negative impact of low-level blast exposure, injuries and symptoms may be underdiagnosed in those exposed to high cumulative low-level blast overpressure.(Elder et al., 2014) Because there is no single event or clinically significant exposure, diagnosis may be more difficult. To date, few studies have been completed investigating the mental health of people primarily exposed to occupational low-level blast exposure. Given the limited research and continued risk faced by people in blast-exposed occupations, more studies are needed to better understand the effects. This study aims to establish whether occupational blast exposure is associated with mental health difficulties.

# Methods:

## Participants:

Any active-duty military member or veteran with past or current experience operating mortars or other artillery was eligible to participate in this study. The primary recruitment strategy used was outreach through an online group for veteran mortarmen. Participants completed an online survey consisting of several forms recording demographic information and screening assessments for mental health disorders. Completion of all forms was required to be included in the study. A demographic overview of surveyed participants is found in Table 1. Recruitment of participants occurred in May 2025, and all respondents were compensated for participation in the study. The University of Pennsylvania Institutional Review Board (IRB) determined this study was exempt under 45 CFR 46.104, as interaction with study participants consisted only of a non-identifiable survey.

**Table 1**

*Demographic Information for Groups*

| Variable | Statistic | Survey | TBI Control | No TBI Control |
|---|---|---|---|---|
| **Age (Years)** | Mean (SD) | 37.5 (13.611) | 39.94 (9.55) | 40.04 (10.08) |
| **Gender** | | | | |
| | Male | 40 | 1125 | 220 |
| | Female | 2 | 139 | 63 |
| **Race** | | | | |
| | White | 35 | 931 | 189 |
| | Black or African-American | 2 | 240 | 78 |
| | Asian | 1 | 18 | 10 |
| | American Indian or Alaska Native | 1 | 10 | 1 |
| | Native Hawaiian or Other Pacific Islander | 0 | 11 | 1 |
| | Unknown | 0 | 54 | 4 |
| **Relationship Status** | | | | |
| | Married/Partnered | 20 | 757 | 161 |
| | Single | 12 | 196 | 55 |
| | Separated | 8 | 305 | 66 |
| | Widowed | 1 | 5 | 1 |
| | Other | 1 | 1 | 0 |
| **Military Branch** | Army | 39 | 844 | 182 |
| | Marines | 3 | 192 | 29 |
| | Air Force | 0 | 121 | 39 |
| | Navy | 0 | 98 | 32 |

| | Other | 0 | 6 | 0 |
|---|---|---|---|---|
| **Military Service (Years)** | Mean (SD) | 8.31 (6.58) | 14.54 (8.91) | 14.11 (9.07) |

## Survey instrument:

The survey consisted of 29 questions recording demographics, military history, blast exposure history, traumatic brain injury history, and reported mental health symptoms. Demographic data collected were age, race/ethnicity, gender, marital status, and employment status. Military history included time served, deployment history, and military pay grade. Blast exposure was measured via self-reported artillery firing data and treatment for blast exposure. mTBI was self-reported.

A battery consisting of the Patient Health Questionnaire-9 (PHQ9),(Ahmadi et al., 2023; Katz et al., 2021) Generalized Anxiety Disorder 7-item scale (GAD7),(Ahmadi et al., 2023) Post Traumatic Stress Disorder Checklist for DSM-5 (PCL5),(Ahmadi et al., 2023; LeardMann et al., 2021) and Alcohol Use Disorders Identification Test - Consumption (AUDIT-C)(Crawford et al., 2013) was used to measure mental health symptoms.

## Exclusion criteria:

Respondents who reported evidence of having experienced a moderate-to-severe TBI (loss of consciousness greater than 30 minutes) were excluded from the study to reflect the control group used in the study, which excluded those with a moderate-to-severe TBI. Associations between repetitive low-level blast exposure are primarily with mTBI rather than moderate-to-severe TBI. (Carr et al., 2017) Respondents were also excluded if they had a history of epilepsy, seizure, stroke, cerebral palsy, brain cancer, brain infection, dementia, or a progressive disease, such as AIDS, as these could result in significant effects unrelated to blast exposure. Of the 50 complete survey responses, 42 were used in the final survey analysis after applying exclusion criteria.

## Control groups:

Two control groups were constructed with data from the Federal Interagency Traumatic Brain Injury Research database(National Institutes of Health and U.S. Department of Defense, n.d.) (FITBIR) collected as part of the CENC 1 longitudinal study conducted by Walker et al.(Walker, 2018; Walker et al., 2016) One control group consisted of veterans returning from Iraq and Afghanistan who had never suffered an mTBI exposure and the other consisted of those who had suffered one or more mTBI exposures.(Walker et al., 2016) Table 1 shows there was significant demographic variation between the three groups. Both control groups excluded personnel who had suffered a moderate or severe TBI, as well as personnel with a history of major neurologic or psychiatric disorders, not including common mental health conditions, such as PTSD, depression, anxiety, or bipolar disorder.(Walker et al., 2016) Two control groups were chosen to help account for the effects of mTBI being unobserved in mortarmen (and therefore not able to be controlled for) on the results. Because some mortarmen experienced an mTBI while others did not, the comparison of mortarmen to a control group of veterans who suffered an mTBI and separately to a control group of veterans who did not suffer an mTBI may bracket the true effect of mTBI. (Rosenbaum, 1987)

# Statistical Analysis:

Survey respondents were compared to control groups with a matched pair study design to balance observed covariates. The control groups were matched to the survey group on the following covariates: age, race, gender, years of military service, military branch, and marital status.

Groups were formed by matching on a propensity score model fitted using logistic regression.(Silber et al., 2001) Absolute standardized difference between covariates was measured to determine the quality of the match. An acceptable match was pre-defined as observed covariates having an absolute standardized difference less than 0.2, with an ideal match holding absolute standardized difference below 0.1.(Silber et al., 2001) To further analyze depressive symptoms while controlling for alcohol use disorder and PTSD, we conducted a separate match including AUDIT-C and PCL-5 as covariates.

After producing an acceptable match, regression models were fitted controlling for covariates and the matched set indicator variables to analyze the estimated effect of belonging to the surveyed group of mortarmen compared to the control groups on measures of mental health after matching.

Data were analyzed using R version 4.5.1,(R Core Team, 2025) using the optmatch(Hansen and Klopfer, 2006) and caret(Kuhn, 2008) packages. Our code can be found at https://osf.io/kwgch/overview?view_only=e41903ee02744463a67b6b3957b4b8c7. (Author, 2026)

# Results:

## Quality of match

All observed covariates had standardized differences less than 0.2 after matching, and the majority of covariates had standardized differences below 0.1. The match, including PCL-5 and AUDIT-C total scores as measured covariates, was slightly less balanced but still maintained absolute standardized differences below 0.2 for all covariates. Figures 1, 2, 3, and 4 are Love plots showing the effect of matching on absolute standardized differences between control groups and the study group.

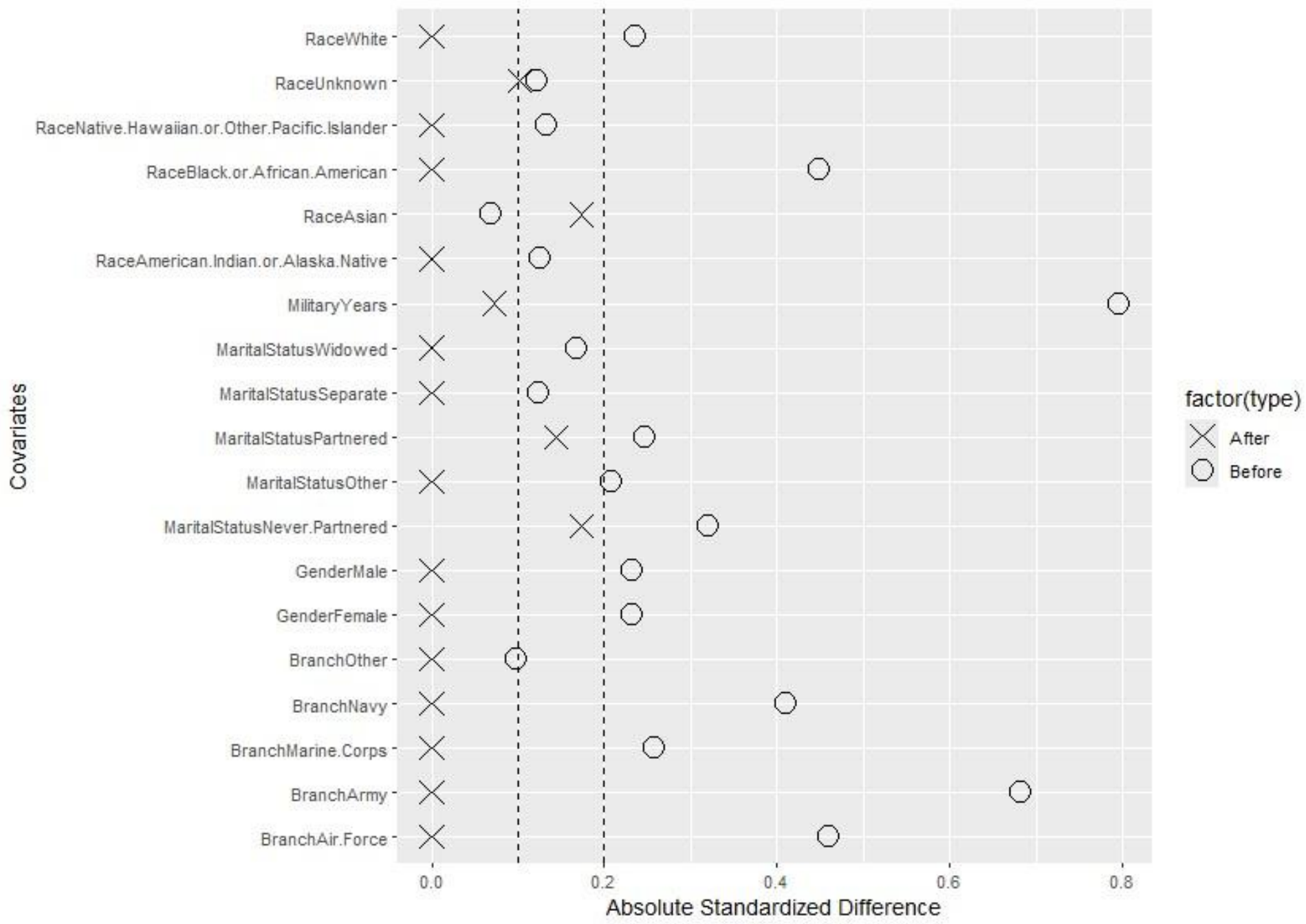

**Figure 1:** Love plot showing the absolute standardized differences in measured covariates between the Survey and TBI control group

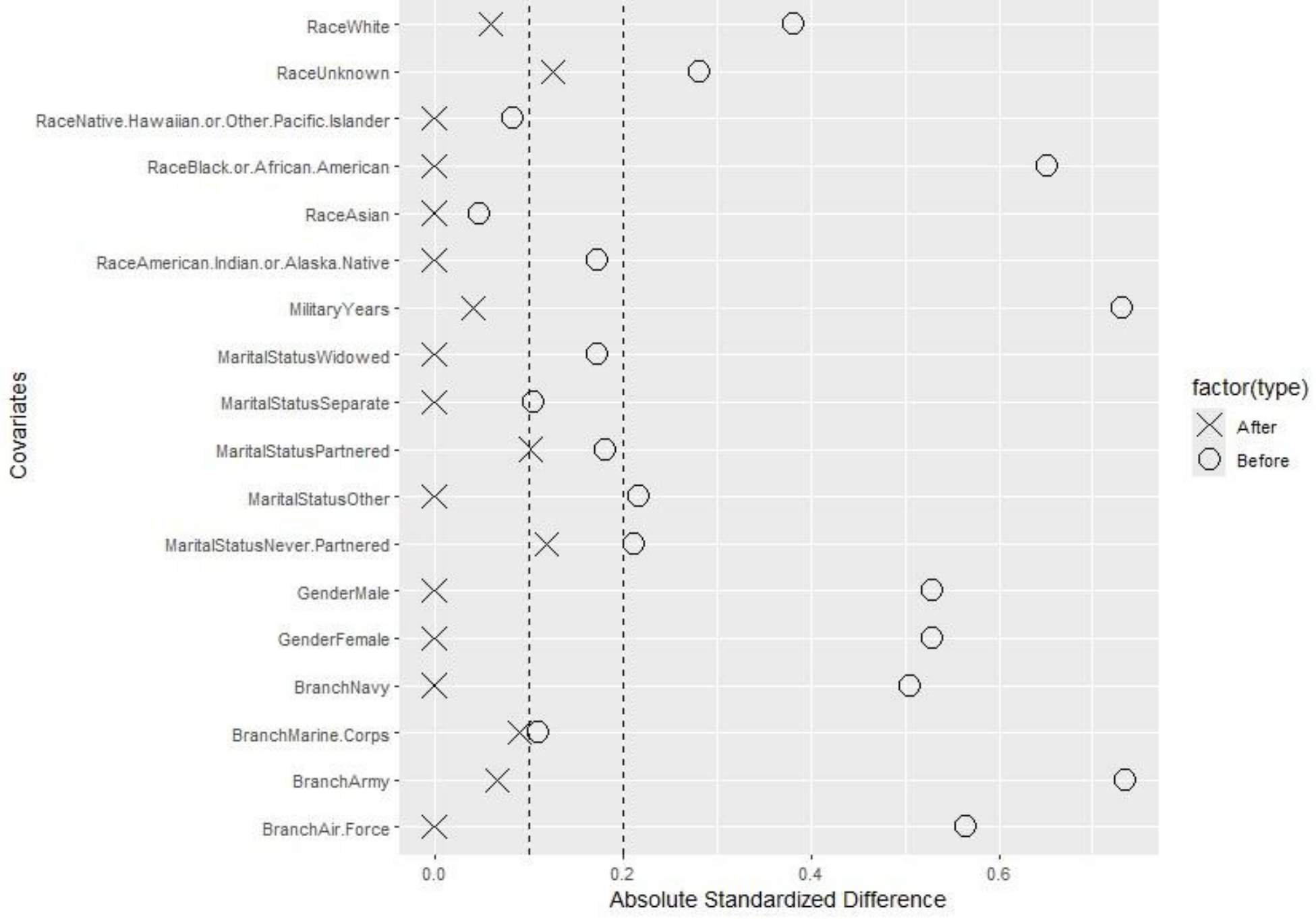

**Figure 2:** Love plot showing the absolute standardized differences in measured covariates between the Survey and No TBI group

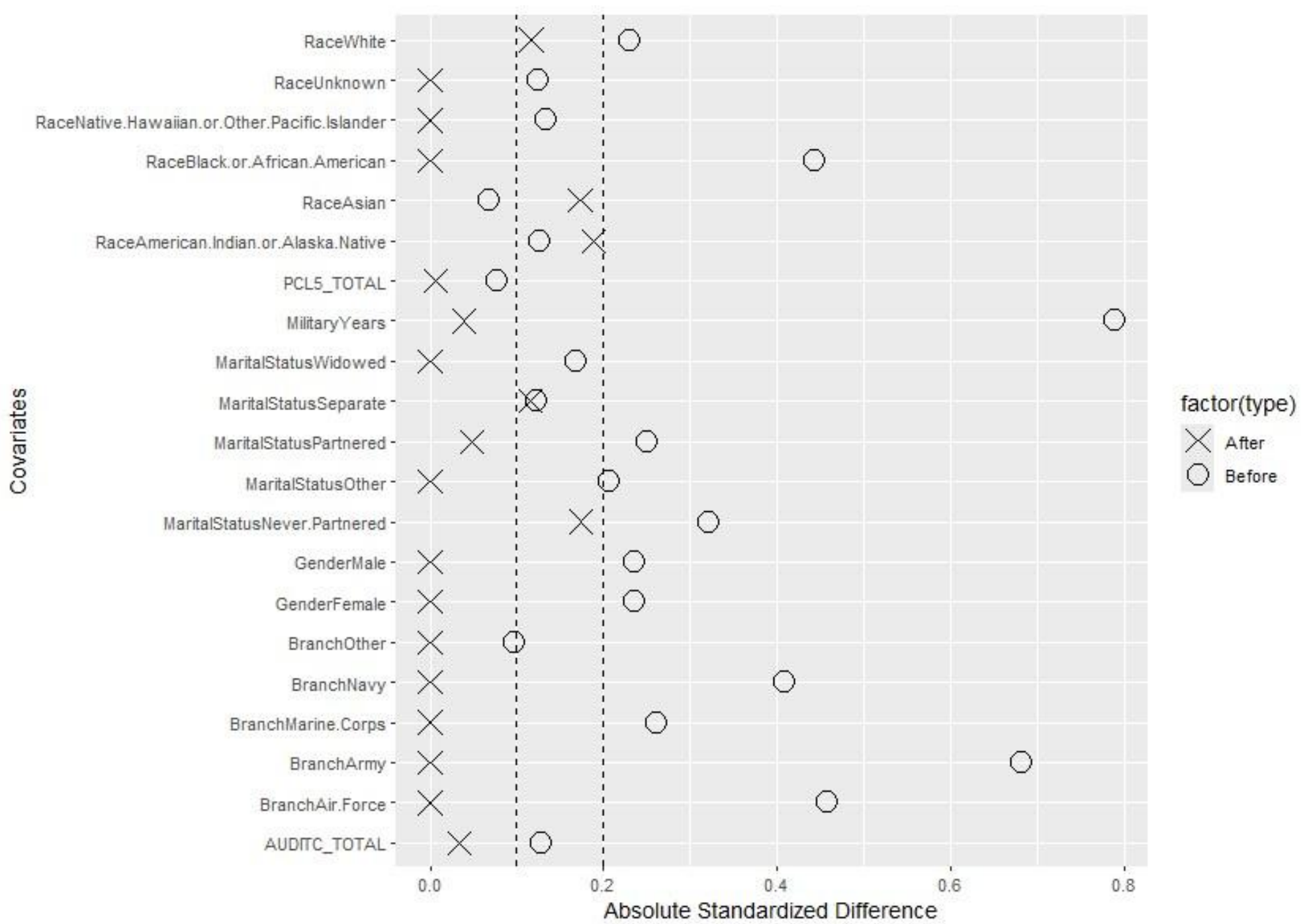

**Figure 3:** Love plot showing the absolute standardized differences in measured covariates, including PCL5 and AUDITC scores, between the Survey and TBI group

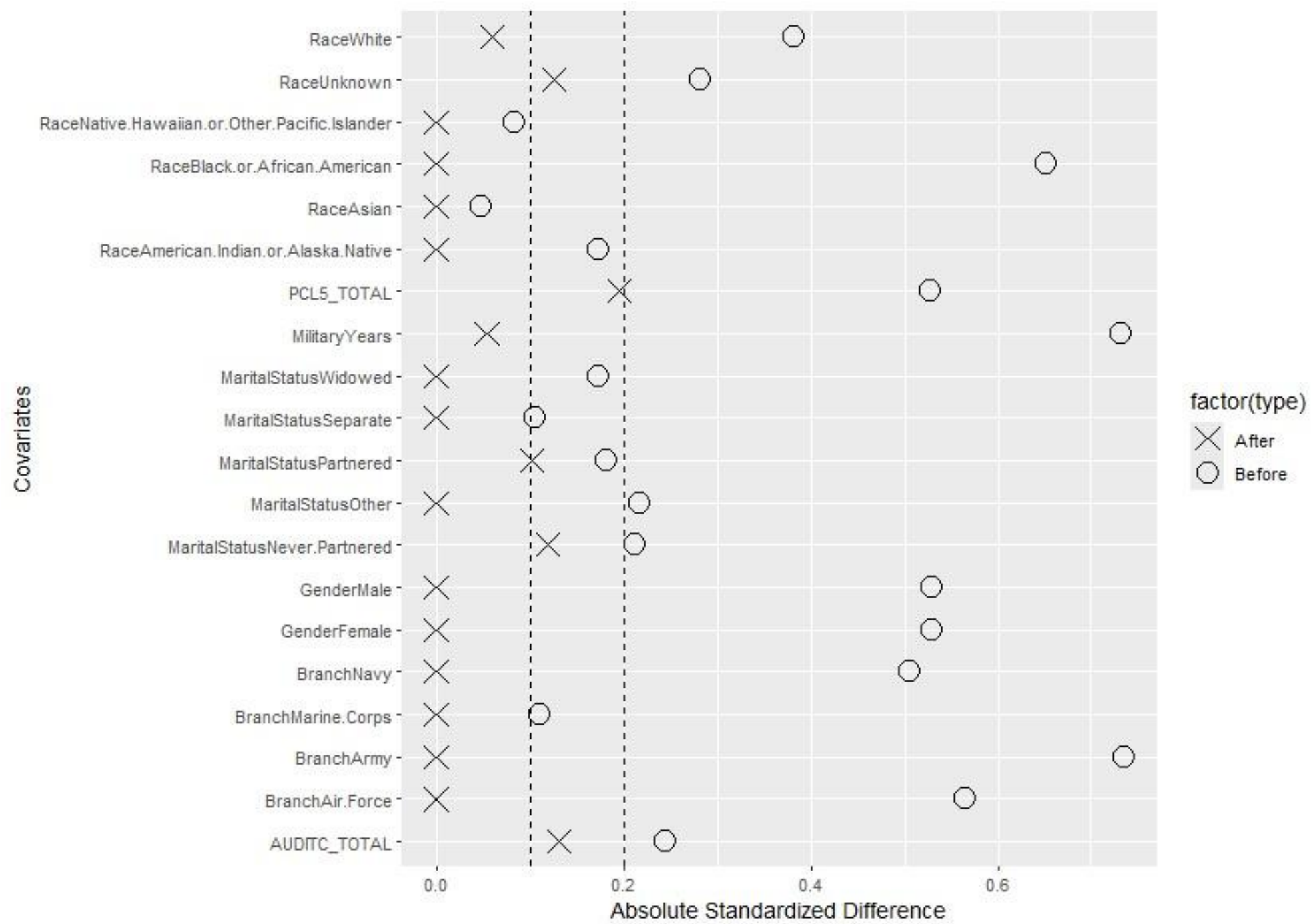

**Figure 4:** Love plot showing the absolute standardized differences in measured covariates, including PCL5 and AUDITC scores, between the Survey and No TBI group

## Association between low-level blast exposure and mental health symptoms:

Table 2 provides descriptive statistics comparing the three groups, showing that mean PHQ-9, PCL-5, and AUDIT-C were all highest in the surveyed group. Compared to the no mTBI control, the survey group of mortarmen had statistically significantly worse measures of depression (estimated difference in PHQ-9 = 7.83, $p < 0.00001$), PTSD (estimated difference in PCL-5 = 16.45, $p < 0.00216$), and alcohol use disorder (estimated difference in AUDIT-C = 1.75, $p < 0.02$). Compared to the mTBI control, no statistically significant difference was found for PTSD or alcohol use disorder. However, the survey group still had statistically significantly worse depression (estimated difference in PHQ-9 = 3.46, $p < 0.04$).

Controlling for PCL-5 and AUDIT-C scores as observed covariates, the surveyed group of mortarmen still had statistically significantly worse depression compared to the no mTBI control (estimated difference in PHQ-9 = 3.74, $p = 0.00097$) and mTBI control (estimated difference in PHQ-9 = 2.68, $p < 0.0035$).

**Table 2**

*Descriptive Statistics for Groups*

| Variable | Statistic | Survey | TBI Control | No TBI Control |
|---|---|---|---|---|
| **PHQ-9 Total Score** | Mean (SD) | 12.43 (6.59) | 8.76 (6.11) | 6.19 (5.63) |
| | No Depression (0-4) (%) | 5 (11.9%) | 370 (29.4%) | 133 (47.2%) |
| | Mild Depression (5-9) (%) | 10 (23.8%) | 371 (29.5%) | 80 (28.4%) |
| | Moderate Depression (10-14) (%) | 12 (28.6%) | 287 (22.8%) | 42 (14.9%) |
| | Moderately Severe Depression (15-19) (%) | 9 (21.4%) | 150 (11.9%) | 17 (6%) |
| | Severe Depression (20-27) (%) | 6 (14.3%) | 79 (6.3%) | 19 (3.5%) |
| **PCL-5 Total Score** | Mean (SD) | 30.88 (21.78) | 29.19 (19.49) | 20.3 (18.1) |
| | n > 33 (%) | 18 (42.9%) | 507 (40.3%) | 68 (24.1%) |
| **AUDIT-C Total Score** | Mean (SD) | 3.29 (3.14) | 2.93 (2.51) | 2.62 (2.25) |

# Discussion:

Veteran mortarmen scored significantly higher on measures of depression, PTSD, and alcohol use disorder compared to veterans who had never suffered an mTBI, and significantly higher on measures of depression compared to veterans who had suffered at least one mTBI exposure. Table 2 shows that 42.9% (n=18) of survey respondents reported PCL-5 scores greater than 33,(Forkus et al., 2023) sufficient for a presumptive diagnosis of PTSD, compared to 24.1% (n=68) of those in the no mTBI control and 40.3% (n=504) in the mTBI control. Despite this, the difference in symptoms of depression could not be accounted for by PCL-5 scores, indicating that the etiology of their depressive symptoms was not entirely PTSD-related. Because PTSD requires a stressful experience, it is possible that the occupation-related effects on mental health facing mortarmen are not measured by a diagnostic for PTSD.(Scott and Stradling, 1994) In addition, the diagnosis for mTBI remains clinical,(Mac Donald et al., 2011; Silverberg et al., 2023) meaning it may be underdiagnosed in people not clearly exposed to normal TBI etiologies.

Another comparison is provided by the population of veterans returning from Iraq surveyed by Hoge (2008), which found that of those who had suffered an injury involving a loss of consciousness, 43.9% had PTSD and 22.9% had major depression using cutoff scores of 10 on PHQ9 for major depressive disorder and 33 on PCL5 for PTSD.(Forkus et al., 2023; He et al., 2020; Hoge et al., 2008) Table 2 shows that the former mortarmen surveyed in this study had a similar reported incidence of PTSD (42.9%) but significantly higher rates of depression (64.3%), providing corroboration of the high rate of depression found in the surveyed population. This comparison is consistent with our findings comparing our sample to FITBIR controls, suggesting that the mortarmen we surveyed were more depressed than similar veterans and that this difference is not accounted for by differences in PTSD. Other previous studies have also found a relationship between a history of chronic blast exposure and symptom severity.(Carr et al., 2015; Stone et al., 2020)

## Limitations:

While we found a significant link between occupational low-level blast exposure and depression, we acknowledge several limiting factors in this study. First, the surveyed group is not necessarily representative of the mortarmen population at large, as the respondents were a convenience sample. It is possible that the survey group overrepresents mortarmen suffering from mental health issues. However, it is also possible that those who are most severely affected by blast-related mental health issues would not be able or available to participate in the survey. Second, both control groups consisted of veterans who had contact with the Department of Veterans Affairs, meaning they were more likely to be receiving some form of treatment, while survey respondents had not necessarily received treatment. Additionally, mTBI is heterogeneous in both etiology and outcome, making it harder to understand its effects on a population and limiting its interpretability.(Rosenbaum and Lipton, 2012) Further, the effects of mTBI and similar injuries may decrease over long periods of time.(Ponsford et al., 2014) This could result in an unobserved covariate of time since injury.

# Conclusion:

Mortarmen suffer from worse mental health symptoms than comparable military peers, both those who had not experienced an mTBI and those who had experienced at least one. Specifically, surveyed mortarmen suffered from significantly more severe symptoms of depression, even after accounting for measures of PTSD and alcohol abuse. If the broader trends observed in this sample are applicable to the wider population of those exposed regularly to repeated low-level blasts, it is likely that people in such occupations are at serious risk of depression and other mental health issues. Our findings extend previous research, indicating that repetitive low-level blast exposure may be responsible for detrimental neurophysiological effects. More emphasis should be placed on occupational safety and limiting cumulative blast overpressure to reduce the negative impact on those regularly exposed.

# Acknowledgements:

We acknowledge the use of the Federal Interagency Traumatic Brain Injury Research Informatics System (FITBIR) and the CENC 1 longitudinal study conducted by Dr. William Walker et. al for the control group data used in this study. Data used in the preparation of this article reside in the Department of Defense (DOD) and National Institutes of Health (NIH)-supported Federal Interagency Traumatic Brain Injury Research Informatics Systems (FITBIR) in Study 263 (DOI: 10.23718/FITBIR/1518855). This manuscript reflects the views of the authors and does not reflect the opinions or views of the DOD or the NIH. We thank Rob Beckman for helpful suggestions for the survey.